\documentclass[onecolumn,showpacs,superscriptaddress]{revtex4}%
\usepackage{graphicx}
\usepackage{amsmath}
\usepackage{amssymb}
\usepackage{verbatim}
\usepackage{float}
\usepackage{amsfonts}
\newtheorem{theorem}{\bf Theorem}[section]

\begin{document}

\title{A Tale of Two Distributions: From Few To Many Vortices In\\
Quasi-Two-Dimensional Bose-Einstein Condensates}

\author{
T. Kolokolnikov$^{1}$, P.G. Kevrekidis$^{2}$, and R. Carretero-Gonz{\'a}lez$^{3}$}

\address{$^{1}$Department of Mathematics and Statistics,
Dalhousie University Halifax, Nova Scotia, B3H3J5, Canada\\
$^{2}$Department of Mathematics and Statistics,
University of Massachusetts, Amherst MA 01003-4515, USA\\
$^{3}$Nonlinear Dynamical Systems Group,
Department of Mathematics and Statistics, and
Computational Science Research Center, San Diego State University, San Diego
CA, 92182-7720, USA}


\keywords{Vortex dynamics, Bose-Einstein condensates, vortex lattices}


\begin{abstract}
Motivated by the recent successes of particle models in capturing the
precession and interactions of vortex structures in quasi-two-dimensional
Bose-Einstein condensates, we revisit the relevant systems of ordinary
differential equations. We consider the number of vortices $N$ as a 
parameter and
explore the prototypical configurations (``ground states'') that arise in the
case of few or many vortices. In the case of few vortices, we modify the
classical result of Havelock [Phil.~Mag.~\textbf{11}, 617 (1931)] illustrating
that vortex polygons in the form of a ring are unstable for $N \geq7$.
Additionally, we reconcile this modification with the recent identification of
symmetry breaking bifurcations for the cases of $N=2,\dots,5$. 
We also briefly discuss the case of a ring of vortices surrounding
a central vortex (so-called $N+1$ configuration). 
We finally examine
the opposite limit of large $N$ and illustrate how a coarse-graining,
continuum approach enables the accurate identification of the radial
distribution of vortices in that limit.
\end{abstract}




\maketitle


\section{Introduction}

The advent of Bose-Einstein condensates (BECs) has offered intriguing twists
to a number of explorations regarding nonlinear waves, their dynamics and
their mutual interactions~\cite{pethick,stringari,emergent}. The realm of point vortices
constitutes a canonical example of this type. Their exploration has been a
fascinating topic, garnering considerable attention starting from the
fundamental contribution of Lord Kelvin~\cite{kel1}, extending to their
critical role in turbulent dynamics proposed by Onsager~\cite{ons} and
reaching up to more recent explorations in a diverse range of fields. The
latter include (but are not limited to) patterns forming in rotating superfluid
$^{4}$He~\cite{yarmchuk}, electron columns confined in Malmberg-Penning
traps~\cite{fajans} and even magnetized, millimeter sized disks rotating at a
liquid-air interface~\cite{whitesides}. Numerous theoretical advances have
also been made by considering the ordinary differential equations (ODEs) describing
the vortex particles. Among them, we briefly note the classical examination of
the stability of vortices in ring formation~\cite{havel}, the consideration of
higher numbers of vortices, e.g., in Ref.~\cite{ziff} and even of asymmetric
equilibria thereof, e.g., in Ref.~\cite{aref0}. Much of this activity has been
summarized, e.g., in the review~\cite{aref1} and the
book~\cite{newton1}. A more recent exposition tying the classical
fluid point vortex problem of Refs.~\cite{aref1,newton1} with the BEC-oriented
motivation of the present work is given in Ref.~\cite{chamoun}.

In the context of BECs, vortices have been one of the focal themes of numerous
investigations which have now been summarized in many
reviews~\cite{donnely,fetter1,fetter2,usmplb,tsubota}. A considerable volume
of associated experimental efforts was concerned with the creation schemes of
such structures~\cite{middel7,middel14,BPA08,freilich10,middel15}, especially
in the form of unit charge vortices. However, there were also efforts to
produce the (dynamically unstable) vortices of higher topological
charge~\cite{middel16} and observe their decay. Furthermore, numerous studies
focused on vortex lattices with a large number of
vortices~\cite{middel13,eng13a,eng13b}. Recently, the significant advancements of
experimental vortex creation and visualization techniques have led to a new
thrust towards the study of small clusters of vortices. While this direction
was initiated early on with the creation of few same-circulation vortices
in Ref.~\cite{middel8} (and their theoretical examination in Ref.~\cite{castindum}), it
has recently ignited considerable interest due to the systematic formation and
observation of counter-circulating vortex
dipoles~\cite{BPA10,freilich10,middel_pra11,Mottonen_PRA_11}, tripoles~\cite{bagn} and also
sets of 2-, 3-, 4- (and more generally controllably few-) vortices~\cite{navar13}.

One of the features that are especially interesting in connection to this BEC
context concerns the relevance and usefulness of the modeling of the vortices
as point particles whose positions are described by 
ODEs. Such an approach has been used in order to not only offer a
detailed quantitative description of the vortex dynamics (in comparison to the
experiment) as in the case of the vortex dipole~\cite{middel_pra11,Mottonen_PRA_11}, but also
as a tool to unveil subtle bifurcation and symmetry breaking phenomena as in
the case of few co-rotating vortices~\cite{navar13} (again corroborating
experimental observations). It is in that light that we consider a deeper
understanding of the features of such ODEs of particular relevance and
importance within this system. On the other hand, in connection to the
classical and intensely studied point-vortex problem, the BEC setting offers
an intriguing twist. Namely, not only should one consider the pairwise
interaction between the vortices, but the phenomenology is significantly
affected by the precessional motion of \textit{each} vortex within the
parabolic external trap confining the BEC atoms. It is the combination of
these two key features that gives rise to numerous unprecedented phenomena,
such as the existence of an equilibrium vortex dipole (with the vortices
located at a suitable distance from the trap center), or the destabilization
of vortex ring formations for small vortex number $N$.

A natural question is the one concerning the applicability of this
point vortex method to BECs and the potential advantages that this approach
may hold in comparison to the well-established approach of the mean-field
so-called Gross-Pitaevskii 
partial differential equation (PDE) model of this setting. While works
combining theoretical observations based on this method and actual
laboratory experiments and comparisons between the 
two~\cite{freilich10,middel_pra11,navar13,Mottonen_PRA_11} strongly suggest the
usefulness of the method, let us add a few more comments in that regard.
In particular, comparisons of the vortex-particle ODEs with the
Gross-Pitaevskii PDE have been given not only for simpler, oscillatory
dynamics, but also for more complex chaotic dynamics; see, 
e.g., Ref.~\cite{tripole} as a recent example. 
Finally, not only cases where only condensate atoms are present
have been considered, but more recently cases with thermal atoms have also
been studied~\cite{proukas}. 
From these works, there is an emerging understanding of the settings where this
point vortex
approximation may be most suitable. In particular, large chemical potentials
make the vortex increasingly more localized and hence its internal structure
progressively less relevant. Furthermore, the quantum pressure term is 
effectively accounted for in the form of these models and should not a priori
pose a problem. Perhaps the most intricate and less controllable aspect is that
of the vortex-sound interactions, which partially depends on the precise form
and the symmetry of the initial conditions; for a relevant, very
recent discussion 
see Ref.~\cite{proukas} (and references therein).  

The present work aims to explore some of these features in the case of
co-rotating vortices (i.e., vortices of only one charge sign). This is the
most typical experimental situation, given that most setups create the
vortices through the imparting of angular momentum to the
system~\cite{fetter2}. More specifically, we intend to examine the two
opposite limits of experimental tractability:\\[-4.0ex]

\begin{itemize}
\item On the one hand, we consider the case of small vortex numbers $N$. In
this case, the canonical expectation is that the vortices will lead to the
formation of a polygonal ring, given the dynamical stability of such a ring.
Here, we will give a systematic stability analysis of the ring formation
generalizing the classical work of Ref.~\cite{havel} (see also
Ref.~\cite{cabral}). However, this will have two
important side conclusions. On the one hand, it will be shown that while the
classical result is that vortices become unstable for $N>7$ in the ring
formation, here due to the precessional term, even the $N=7$ case is
\textit{always} unstable, non-trivially modifying the classical result.
However, there are more surprises; we will see that the eigenvalues of the
ring formation critically depend on the ring radius and may even lead to
instability for the cases $N=2$ to $N=6$, whereas the work of Ref.~\cite{havel}
predicts generic stability for the classical point-vortex problem. 
In fact, it will be argued that these
destabilization events are exactly the ones recently identified by varying the
angular momentum of the vortex system in Ref.~\cite{navar13}.
\smallskip

\item On the other hand, the opposite limit of large vortex number is equally
interesting (and experimentally accessible, per the vortex lattice experiments
discussed above). In that case, we present a coarse graining description
developing a continuum model for the vortex distribution and its stationary
form. We identify the radial form of this distribution. By finding the stable
equilibrium of the ODEs for the case of large $N$ and developing a vortex
counting algorithm that enables the identification (from the particle results)
of this distribution, we obtain excellent agreement with the prediction of the
coarse grained model.
\end{itemize}

\vspace{-0.2cm}
We believe that these findings will shed light on the theoretical analysis of
vortices in quasi-two-dimensional BECs, identifying some of the complications
and subtleties arising due to the presence of the (critical for the present
dynamics) precessional term. As an aside, we also show how some of the
relevant techniques can be generalized in other settings, e.g., by computing 
the stability of the so-called $N+1$ configuration, where $N$
vortices form a ring, while $1$ is located at the trap center.
It should be noted here that 
a principal motivation of this work concerns settings 
other than the ones where the rotation
of the entire condensate has rendered a multi-vortex state the ground
state of the system (minimizing its free energy due to the presence
of the vortices). More specifically, instead, 
we envision a situation such as that
of Refs.~\cite{BPA10,freilich10,middel_pra11,Mottonen_PRA_11} or Ref.~\cite{navar13} where multiple
vortices have been created through a suitable external driving or
quenching of the condensate, yet they represent an excited,
potentially dynamically stable state of the system.

Our presentation is structured as follows. In Sec.~\ref{sec:setup}, 
we offer the general
theoretical setup in line with the earlier works (such
as Refs.~\cite{middel_pra11,navar13}) that corroborated its correspondence with
experimental results. In Sec.~\ref{sec:ring}, 
we present the analysis of the polygonal
vortex ring and how it connects to the stability conclusions of recent few
vortex studies such as Refs.~\cite{navar13,zampetaki}. 
In Sec.~\ref{sec:continuum}, we explore the
opposite limit of large $N$ and the corresponding coarse-grained, continuum
model. Again a comparison is offered, this time with numerical computations
within that limit. Finally, in Sec.~\ref{sec:conclu}, 
we summarize our findings and
present a number of conclusions and possible future directions. The appendices
present a number of technical details associated, e.g., with the stability of
the $N+1$ vortex configuration.

\section{Theoretical Setup} 
\label{sec:setup}

The starting point for our considerations will consist of the vortex equation
of motion which can be written in the form of a single complex-valued
ODE for $z_{j}=X_{j} + i Y_{j}$, where $(X_{j},Y_{j})$ denotes the planar
position of the $j$-th vortex. This equation reads:
\begin{equation}
\dot{z}_{j}=if\left(  \left\vert z_{j}\right\vert \right)  z_{j}+ic\sum_{k\neq
j}\frac{z_{j}-z_{k}}{\left\vert z_{j}-z_{k}\right\vert ^{2}},\ \ \ j=1\ldots
N. \label{bec}%
\end{equation}
In the right hand side of Eq.~(\ref{bec}) the first term represents the
precession of the vortex around the trap center. Here we consider only
vortices of a unit charge (given their dynamical stability) and assume that
all have the same charge (without loss of generality we assume this to be
$S=+1$), as in the context of the recent experiments of Ref.~\cite{navar13}. While
for many of our considerations, we will keep this precession term as general
as possible, for a number of concrete calculations we will assume the same
form as used in the recent experimental considerations
of Refs.~\cite{middel_pra11,navar13}; see also the detailed analysis/comparison with
single vortex precession experiments in Ref.~\cite{freilich10}. In particular, in
line with these works, we will choose:
\begin{equation}
f(r)=\frac{a}{1-r^{2}}. \label{f}%
\end{equation}
Here, the radius has been normalized to the Thomas-Fermi radius (roughly
tantamount to the spatial extent of the BEC), while the factor $a$
representing the precession frequency of the vortex very near the center of
the trap can be absorbed into a rescaling of time (rendering time
dimensionless).

On the other hand, the second term in Eq.~(\ref{bec}) represents the
vortex-vortex interaction, i.e., the velocity field at the location of a vortex
due to the presence of other surrounding vortices. Our adimensionalization of
the model follows that of Ref.~\cite{navar13}, where it was explained that a
``typical'' experimentally relevant value for
$c$ is $0.1$. It should be noted here that this constant is effectively taking
into account the non-uniformity of the background density (due to the presence
of the trap) through which the vortices are interacting. However, one can
consider more elaborate (integral) functional forms, explicitly taking into
account the density modulation, as described, e.g., in Ref.~\cite{busch}. Given the
successes of the simpler set up in capturing the recent experimental
observations and the amenability of the corresponding functional forms to our
analytical considerations, we will indeed proceed to consider the precession
and interaction terms as presented in Eq.~(\ref{bec}).

It is natural to expect in the considered case of co-rotating vortices that
both the precession and interaction effects lead to rotation of the vortices
in the same direction. In that light, no genuine equilibria may exist but
instead we may seek relative (in a rotating frame) equilibria of the 
form $z_{j}(t)=e^{i\omega t}x_{j}$. These satisfy the equation
\begin{equation}
\omega x_{j}=f\left(  \left\vert x_{j}\right\vert \right)  x_{j}+c\sum_{k\neq
j}\frac{x_{j}-x_{k}}{\left\vert x_{j}-x_{k}\right\vert ^{2}},
\end{equation}
which will be the main focus of our considerations in what follows.

As a (partially numerical) aside, we will also consider the following
``aggregation'' equation:
\begin{equation}
\dot{x}_{j}=\left(  f(\left\vert x_{j}\right\vert )-\omega\right)  x_{j}%
+c\sum_{k\neq j}\frac{x_{j}-x_{k}}{\left\vert x_{j}-x_{k}\right\vert ^{2}}.
\label{aggr}%
\end{equation}
By construction, the relative equilibrium $z_{j}(t)=e^{i\omega t}x_{j}$ of
Eq.~(\ref{bec}) corresponds to the relaxational equilibrium $x_{j}(t)=x_{j}$ of
Eq.~(\ref{aggr})\ and vice-versa. Moreover there is an intimate connection between
the stability of the two models. In Appendix \ref{appI} we prove the following result.

\begin{theorem}
\label{thm:stab}Suppose that an equilibrium $x_{j}(t)=\xi_{j}$ of Eq.~(\ref{aggr})
is stable. Then the relative equilibrium $z_{j}(t)=e^{i\omega t}\xi_{j}$ of
Eq.~(\ref{bec}) is (neutrally)\ stable. The converse is also true in the following
two cases:\ (i) $f\left(  r\right)  =\operatorname{const.}$ or
(ii)\ $\left\vert \xi_{j}\right\vert =\operatorname{const.}$ for all $j.$
\end{theorem}

The case (i) of this theorem was shown in Ref.~\cite{Kolokolnikov:2013}; this is
reproduced in Appendix \ref{appI}. However the proof in Ref.~\cite{Kolokolnikov:2013}
does not work for a general case
of non-homogeneous $f(r)$, and a more general approach is taken here. On the
flip side, we can only show the stability of Eq.~(\ref{aggr}) implies stability of
Eq.~(\ref{bec}); we do not know how to prove the converse (nor do we have counter-examples).

\section{Ring solutions}
\label{sec:ring}

The prototypical configuration that one expects to identify as a stable
equilibrium in the case of small vortex number $N$ (motivated by the
corresponding result in the absence of precession~\cite{havel,ziff}) is the
``ring'' configuration with the vortices sitting at the vertices of a
canonical polygon.
Assuming such a relative equilibrium to Eq.~(\ref{bec}) with radius $R$, we
can write it in our complex notation as:
\begin{equation}
z_{j}(t)=R\exp\left(  i\omega t\right)  \exp\left(  \frac{2\pi i}{N}j\right)
. \label{ringAnzatz}%
\end{equation}
In this case, we can compute:
\begin{equation}
\sum_{k\neq j}\frac{z_{j}-z_{k}}{\left\vert z_{j}-z_{k}\right\vert ^{2}}
=\frac{1}{R}\exp\left(  i\omega t+\frac{2\pi i}{N}j\right)
\frac{N-1}{2},
\end{equation}
where we have used the identity%
\[
\sum_{k=1}^{N-1}\frac{1}{1-\exp\left(  -\frac{2\pi i}{N}k\right)  }=\frac{N-1}{2}.
\]
Therefore the radius $R$ satisfies:
\begin{equation}
\omega=f(R)+\frac{c(N-1)}{2R^{2}}. \label{Rss}%
\end{equation}

As the case (ii) of Theorem \ref{thm:stab} [of Appendix \ref{appI}] shows, the ring for the vortex model
Eq.~(\ref{bec})\ is stable if and only if it is stable for the aggregation model
Eq.~(\ref{aggr}). The full characterization of linear stability is given by the
following theorem (following the procedure used, e.g.,
in Refs.~\cite{Kolokolnikov:2010,bigring}).


\begin{theorem}
\label{thm:stabring} Consider the ring solution for  Eq.~(\ref{bec}), of radius $R$
as given by Eq.~(\ref{ringAnzatz}), where the frequency $\omega$ is given by
Eq.~(\ref{Rss}). Suppose that $N$ is odd. Then the ring is stable provided that
\[
f^{\prime}(R)R+\frac{c}{8R^{2}}\left(  N-1\right)  (N-7)<0,
\]
and is unstable if the inequality is reversed. Suppose that $N$ is even. Then
the ring is stable provided that
\[
f^{\prime}(R)R+\frac{c}{8R^{2}}\left(  N^{2}-8N+8\right)  <0,
\]
and is unstable if the inequality is reversed.

The ring is generically unstable if $N\geq7$ and $f(r)$ is an increasing function.
\end{theorem}

\textbf{Proof.} Because of Theorem \ref{thm:stab} case (ii), it is sufficient to
consider the stability of the steady state $x_{k}(t)=R\exp\left(  2\pi
ik/N\right)  $ of the aggregation equation (\ref{aggr}). Consider small
perturbations of this state, of the form%
\[
x_{k}(t)=R\exp\left(\frac{ 2\pi ik}{N}\right)  \left(  1+h_{k}(t)\right)
,\ \ \ |h_{k}|\ll 1,
\]
where $h_k$ is a small, complex-valued, perturbation.
After some algebra we obtain for the evolution of the small perturbations
\begin{align}
\frac{dh_{j}}{dt}=\left(  f^{\prime}(R)\frac{R}{2}+f(R)-\omega\right)
h_{j}+f^{\prime}(R)\frac{R}{2}\bar{h}_{j}
+ c\sum_{k\neq j}\frac{\exp\left(
2\pi\left(  k-j\right)  /N\right)  \bar{h}_{j}-\bar{h}_{k}}{4R^{2}\sin
^{2}\left(  \frac{\pi\left(  k-j\right)  }{N}\right)  }, 
\label{h}%
\end{align}
where the overbar denotes complex conjugation.
Using the following Fourier mode de\-com\-po\-si\-tion for the perturbation:
\begin{equation}
h_{j}(t)=\xi_{+}(t)e^{im2\pi j/N}+\xi_{-}(t)e^{-im2\pi j/N},\ \ \
m\in\mathbb{N}, \label{29july1:04}
\end{equation}
and collecting like terms in $e^{im2\pi j/N}$ and $e^{-im2\pi j/N},$ the
system (\ref{h})\ decouples into a sequence of $2 \times2$ subproblems
\begin{align}
\xi_{+}^{\prime}&=
\left(  f^{\prime}(R)\frac{R}{2}+f(R)-\omega\right)
\xi_{+}+f^{\prime}(R)\frac{R}{2}\bar{\xi}_{-}
+\sigma_+\bar{\xi}_{-},
\label{redflower}
\\[2.0ex]
\xi_{-}^{\prime}&=
\left(  f^{\prime}(R)\frac{R}{2}+f(R)-\omega\right)
\xi_{-}+f^{\prime}(R)\frac{R}{2}\bar{\xi}_{+}
+\sigma_-\bar{\xi}_{+},
\label{whiteflower}
\end{align}
where
\begin{align}
\sigma_+
&\equiv
c\sum_{k,k\neq
j}\frac{e^{i2\pi\left(  k-j\right)  /N}-e^{im2\pi\left(  k-j\right)  /N}%
}{4R^{2}\sin^{2}\left(  \pi\left(  k-j\right)  /N\right)  },
\nonumber
\\
\nonumber
\sigma_-
&\equiv
c\sum_{k,k\neq
j}\frac{e^{i2\pi\left(  k-j\right)  /N}-e^{-im2\pi\left(  k-j\right)  /N}%
}{4R^{2}\sin^{2}\left(  \pi\left(  k-j\right)  /N\right)  }.
\end{align}
Using the following identity:
%
\begin{equation}
\sum_{k=1}^{N-1}\frac{e^{\pm im2\pi k/N}}{\sin^{2}\left(  \pi k/N\right)
}=2\left(  m-\frac{N}{2}\right)  ^{2}-\frac{N^{2}}{6}-\frac{1}{3},\ \ m=0\ldots N,
\end{equation}
it is possible to write
\begin{equation}
\sigma \equiv
\sigma_+ =
\sigma_- =
\frac{c}{2R^{2}}\left(  m-1\right)  \left(N-m-1\right).
\end{equation}
Taking the conjugate of Eq.~(\ref{whiteflower}), the system can be written as
\begin{equation}
\left(
\begin{array}
[c]{c}%
\xi_{+}^{\prime}\\[2.0ex]
\bar{\xi}_{-}^{\prime}%
\end{array}
\right)  =\left(
\begin{array}
[c]{cc}%
f^{\prime}(R)\frac{R}{2}+f(R)-\omega
& f^{\prime}(R)\frac{R}{2}+
\sigma
\\[2.0ex]
f^{\prime}(R)\frac{R}{2}+
\sigma
&
f^{\prime}(R)\frac{R}{2}+f(R)-\omega
\end{array}
\right)  \left(
\begin{array}
[c]{c}%
\xi_{+}\\[2.0ex]
\bar{\xi}_{-}%
\end{array}
\right),  \label{gulag}%
\end{equation}
whose eigenvalues are given by%
\[
\lambda_{\pm}(m)=f^{\prime}(R)\frac{R}{2}+f(R)-\omega\pm\left(  f^{\prime
}(R)\frac{R}{2}+\frac{c}{2R^{2}}\left(  m-1\right)  \left(  N-m-1\right)
\right)  ,\ \ m=0\ldots N-1.
\]
Using Eq.~(\ref{Rss})\ this simplifies to
\begin{equation}
\begin{array}{rcl}
\lambda_{-}(m)&=&
\displaystyle
\frac{c}{2R^{2}}\left[  -(N-1)-\left(  m-1\right)  \left(
N-m-1\right)  \right],
\\[2.0ex]
\lambda_{+}(m)&=&
\displaystyle
f^{\prime}(R)R+\frac{c}{2R^{2}}\left[  \left(  m-1\right)
\left(  N-m-1\right)  -(N-1)\right]  . \label{lamp}%
\end{array}
\end{equation}
Note that $\lambda_{-}(0)=0$; this mode corresponds to rotation invariance.
Moreover $\left(  m-1\right)  \left(  N-m-1\right)  $ is positive for all
integers $1\leq m\leq N-1$ and attains its max at $m=N/2.$ Therefore
$\lambda_{-}(m)$ correspond to stable eigendirections and the instability
threshold is obtained by setting $\lambda_{+}(\lfloor N/2 \rfloor)=0$, 
where $\lfloor\cdot\rfloor$ denotes the floor function.

Suppose that $N$ is odd. Then we substitute $m=(N-1)/2$ into Eq.~(\ref{lamp}) to
obtain%
\[
\lambda_{+}((N-1)/2)=f^{\prime}(R)R+\frac{c}{8R^{2}}\left(  N-1\right)
(N-7),\ \ \ N\text{ odd}.%
\]
If $N$ is even we substitute $m=N/2$ which yields%
\[
\lambda_{+}(N/2)=f^{\prime}(R)R+\frac{c}{8R^{2}}\left(  N^{2}-8N+8\right)
,\ \ \ N\text{ even.}%
\]
The ring is stable for the aggregation equation (and by extension for the
actual vortex model) if the right hand side of the above two equations
(respectively, for $N$ odd and even) is negative and unstable otherwise. If
$f^{\prime}=0,$ this recovers the $N=7$ threshold. $\ \blacksquare$

Some observations are important to make here. In particular, we wish to
examine the special cases of $N=2, \dots, 7$ which are well-known to be canonical
examples where the polygonal ring is of interest as a configuration even in
the absence of precession.

We start with the $N=7$ case. We can see that contrary to the
case where the precession is absent (wherein this case is \textit{critical}
with the corresponding eigenvalue being neutral), here the presence of
precession has a crucial impact. In particular, it adds a positive part [for
$f(R)$ increasing, as is the case in our BECs] to the pertinent eigenvalue
rendering the corresponding configuration \textit{generically} unstable. This
is a particular trait of our BEC vortex ring configuration.

For the cases with $N<7$, we can express the corresponding eigenvalue in the
general form $\lambda=f^{\prime}(R)R+\frac{c}{8R^{2}}\left(  N^{2}%
-8N+s\right)  $, where $s=7$ for $N$ odd and $s=8$ for $N$ even. A key
observation now (which directly connects the present work with the
experimental observations of Ref.~\cite{navar13} and the computational/theoretical
analysis of Ref.~\cite{zampetaki}) is that although for these $N<7$ cases the
configuration is \textit{not} generically unstable, nevertheless it 
\textit{may}
be unstable for sufficiently large $R$. To phrase this differently, we can
parametrize the vortex system by the angular momentum of the vortex particles,
which is a conserved quantity of the dynamics. The angular momentum is defined
as $L=\sum_{j=1}^{N}|z_{j}|^{2}$ and in the case of the ring configuration
acquires the especially simple form $L=NR^{2}$.%
\footnote{
It is worth noting here that while this is the standard form for
the angular momentum in the context of point vortices, cf.~Eq.~(2.13) in 
p.~514 of the review~\cite{chamoun}, this is {\it not identical} to the
angular momentum of the Bose-Einstein condensate, which is well-established,
e.g., in the context of the corresponding PDE
model of the Gross-Pitaevskii equation~\cite{pethick,stringari,emergent}. 
In our considerations here, we will use the term for the former
and not the one for the latter.
}
In that light and taking
$f(r)$ as in Eq.~(\ref{f}) with $a=1$, the eigenvalue whose zero crossing will
determine the potential instability of a configuration with $N=2,\dots,6$
reads:
\[
\lambda=\frac{2L}{N(1-\frac{L}{N})^{2}}+\frac{cN}{8L}\left(  N^{2}%
-8N+s\right).
\]
It is then straightforward to infer that the critical angular momentum given
by setting $\lambda=0$ satisfies
\[
16L^{2}+c(N-L)^{2}(N^{2}-8N+s)=0
\]
which yields a critical $L$ of the form:
\[
L_{cr}=\frac{cN(N^{2}-8N+s)+4N\sqrt{{c(8N-N^{2}-s)}}}{16+c(N^{2}-8N+s)}.%
\]

Remarkably, this expression yields the relevant critical angular momenta for
\textit{all} cases of $N=2$ to $N=5$ in direct agreement with the expressions
given in Refs.~\cite{navar13,zampetaki} for the symmetry breaking bifurcations due
to the destabilization of the ring configuration. Using the notation
\begin{align}
r_{1}^{2}\equiv\frac{\sqrt{c}}{\sqrt{c}+2}, 
\quad 
r_{2}^{2}\equiv\frac{\sqrt{c}}{\sqrt{c}+ \sqrt{2}}, 
\label{lextra4}%
\end{align}
we have that:
\begin{align}
\nonumber
L_{cr,N=2}=2\, r_{1}^{2},
\\
\nonumber
L_{cr,N=3}=3\, r_{2}^{2},
\\
L_{cr,N=4}=4\, r_{2}^{2},
\\
\nonumber
L_{cr,N=5}=5\, r_{2}^{2},
\\
\nonumber
L_{cr,N=6}=6\, r_{1}^{2}.
\label{lextra5}%
\end{align}
It is rather intriguing that the critical angular momentum for the different
cases exhibits an apparent pattern although not a clearly definite one.
Interestingly, even for $N=7$, while the dominant eigenvalue is always
positive (as indicated above), even the next one $\lambda_{(N-3)/2}$ can be
seen to cross $0$ at $L_{cr,N=7}=7\,r_{2}^{2}$, extending this interesting
pattern. We also note in passing that for the case of $N=6$, the relevant
critical point had not been previously identified in Ref.~\cite{navar13}
or Ref.~\cite{zampetaki}.

However, a final observation is also in order. As discussed
in Ref.~\cite{zampetaki}, the interval (of $L$) of dynamical stability of the small
$N$ configurations does \textit{not} coincide with the interval of $L$ for
which these constitute the ground state of the system. In the case of $N=2$,
the stability threshold and ground state asymmetry threshold coincide (this is
a supercritical pitchfork point), but in other cases such as $N=3$ and $N=4$,
asymmetric configurations (such as isosceles triangles for $N=3$ or rhombic
configurations for $N=4$) acquire lower energy than the polygonal ring
distinctly before its loss of stability threshold~\cite{zampetaki}. 
Namely, the ring
configuration becomes a local (rather than global) minimum of the energy
clearly before its destabilization. Unfortunately, these asymmetric
configurations which are stabilized by the presence of our precession term (in
its absence such asymmetric configurations are unstable, as discussed, e.g.,
in Ref.~\cite{aref0}), do not have a general closed form that would enable their
stability analysis. Nevertheless, another symmetric configuration that emerges
as relevant for the ground state of the system, at least in the case of $N=5$
examined in Ref.~\cite{zampetaki} (and obviously also for larger $N$) is the
so-called $N+1$ vortex configuration, consisting of $N$ vortices on the
polygonal ring and one more at the center. For the classical vortex problem
[$f=0$ in Eq.~(\ref{bec})], the stability of this configuration was analyzed in
Ref.~\cite{cabral2000stability}; see also Ref.~\cite{barry}.
In Appendix \ref{appII} we show the
following generalization to the full BEC problem:

\begin{theorem}
\label{thm:N+1}Consider the $N+1$ configuration of Eq.~(\ref{bec})\ consisting of
$N$ vortices uniformly distributed on a ring of radius $R$ and angular
velocity $\omega$ given by Eq.~(\ref{ringAnzatz})\ plus a vortex at the origin.
Then $R$ satisfies%
\begin{equation}
\omega=f(R)+\frac{c}{2R^{2}}\left(  N+1\right)  . \label{R2}%
\end{equation}
Let
\begin{equation}
\lambda_{+}^{\ast}(m)=\lambda_{+}(m)-2c/R^{2} \label{lampstar}%
\end{equation}
where $\lambda_{+}(m)$ as given by Eq.~(\ref{lamp})\ and let
\begin{equation}
\label{M0}
M_{0}=\left(
\begin{array}
[c]{ccc}%
f^{\prime}(R)\frac{R}{2}-\frac{c\left(  N+1\right)  }{2R^{2}} & f^{\prime
}(R)\frac{R}{2}-\frac{c}{R^{2}} & \frac{c}{R^{2}}
\\[2.0ex]
f^{\prime}(R)\frac{R}{2}-\frac{c}{R^{2}} & f^{\prime}(R)\frac{R}{2}%
-\frac{c\left(  N+1\right)  }{2R^{2}} & 0
\\[2.0ex]
\frac{cN}{R^{2}} & 0 &  f(0)-f(R)-\frac{c\left(  N+1\right)  }{2R^{2}}
\end{array}
\right)  .
\end{equation}
The $N+1$ configuration is stable if and only if \ $\lambda_{+}^{\ast} \left(
\left\lfloor N/2\right\rfloor \right)  <0$ and all eigenvalues of $M_{0}$ are negative.
\end{theorem}

In particular, this theorem shows that the $N+1$ configuration is stable if
the $N$-ring is stable and if in addition the matrix $M_{0}$ is negative
definite.

When $f=0,$ both $9+1$ and $3+1$ configurations are marginally stable [the
former due to $\lambda_{+}^{\ast}(4)=0$ when $N=9,$ the latter due to the
eigenvalue of $M_{0}$ crossing zero when $N=3$]; the $N+1$ configuration is
stable for $3<N<7$ and is unstable for $N>9$ or $N<3.$ This is in agreement
with the results in Ref.~\cite{cabral2000stability}. When $f$ is increasing as in
Eq.~(\ref{f}), both $9+1$ and $3+1$ configurations lose their marginal stability
and become unstable, so that $N+1$ configuration becomes unstable for any
$N\geq9$ or $N\leq3$.


\section{Continuum limit}
\label{sec:continuum}

Having considered the case of small $N$, we now explore the opposite limit of
large $N$. Notice, however, that our results for the ring and the $N+1$ vortex
configuration are entirely general and the dynamical stability thereof is
obtained for any $N$. Yet, these configurations can only be stable when $N$ is
sufficiently small, as discussed above, e.g., $N<7$ for the ring state (and even
then for sufficiently small $L$). Hence, in the opposite limit of large $N$,
we expect that a substantially different vortex distribution will arise. This
expectation is confirmed not only by the well-known vortex lattice
observations of, e.g., Refs.~\cite{middel13,eng13a,eng13b}, but also by the 
direct numerical evolution of the aggregation equation (\ref{aggr}); see the
top left panel of Fig.~\ref{tkfig1}. Recall that the aggregation equation has
the benefit of relaxing to equilibrium attractors corresponding to the
marginally stable equilibria of our original Eq.~(\ref{bec}). It is then
particularly relevant to attempt to identify the distribution of such a large
number of vortices $N$. Here, we will use techniques similar to Ref.~\cite{fetecau}
to derive the limiting density profile.

\begin{empty}
\begin{figure}[tb]
\begin{tabular}{cc}
\includegraphics[width=0.43\textwidth]{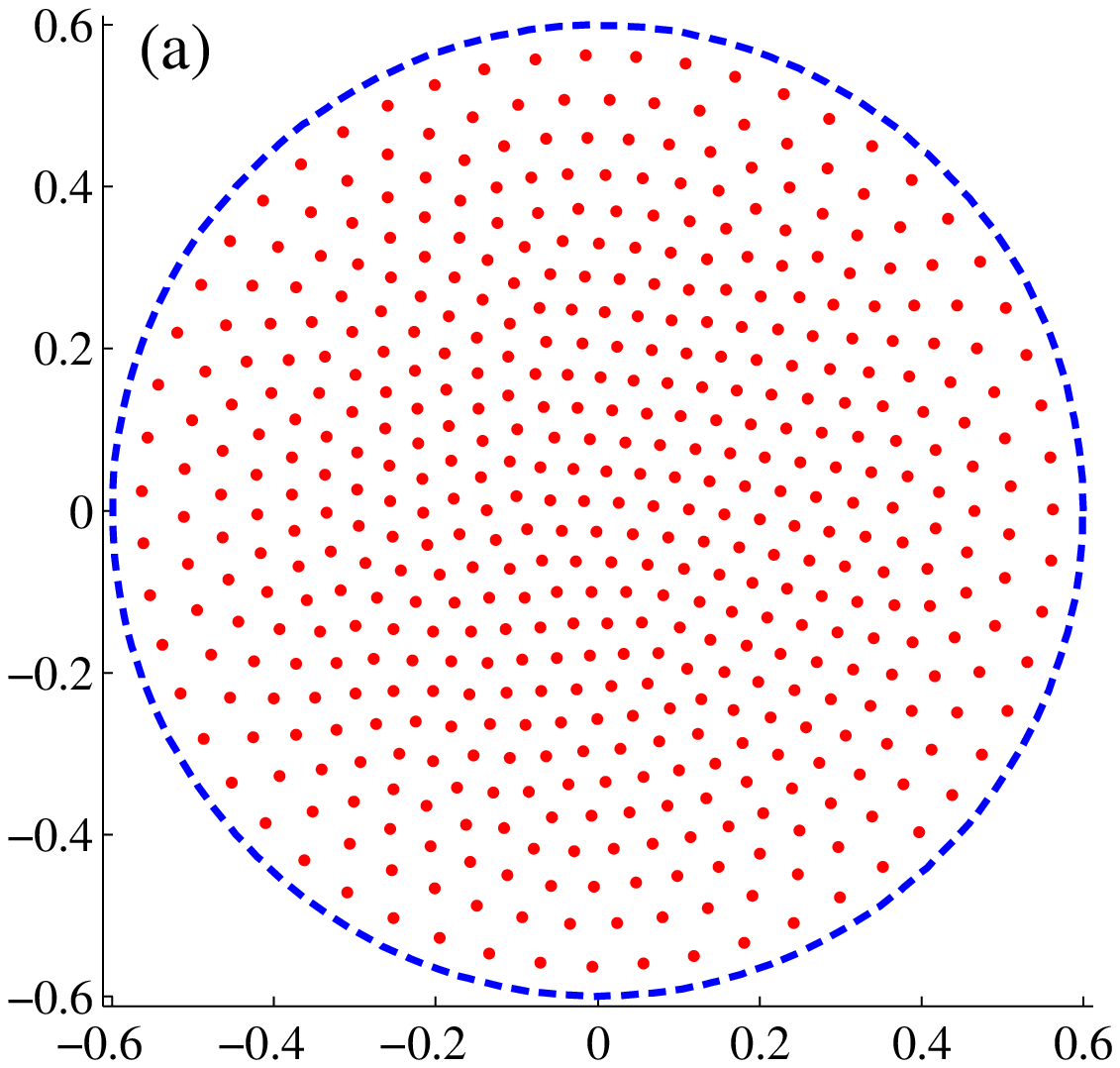}&~~~
\includegraphics[width=0.43\textwidth]{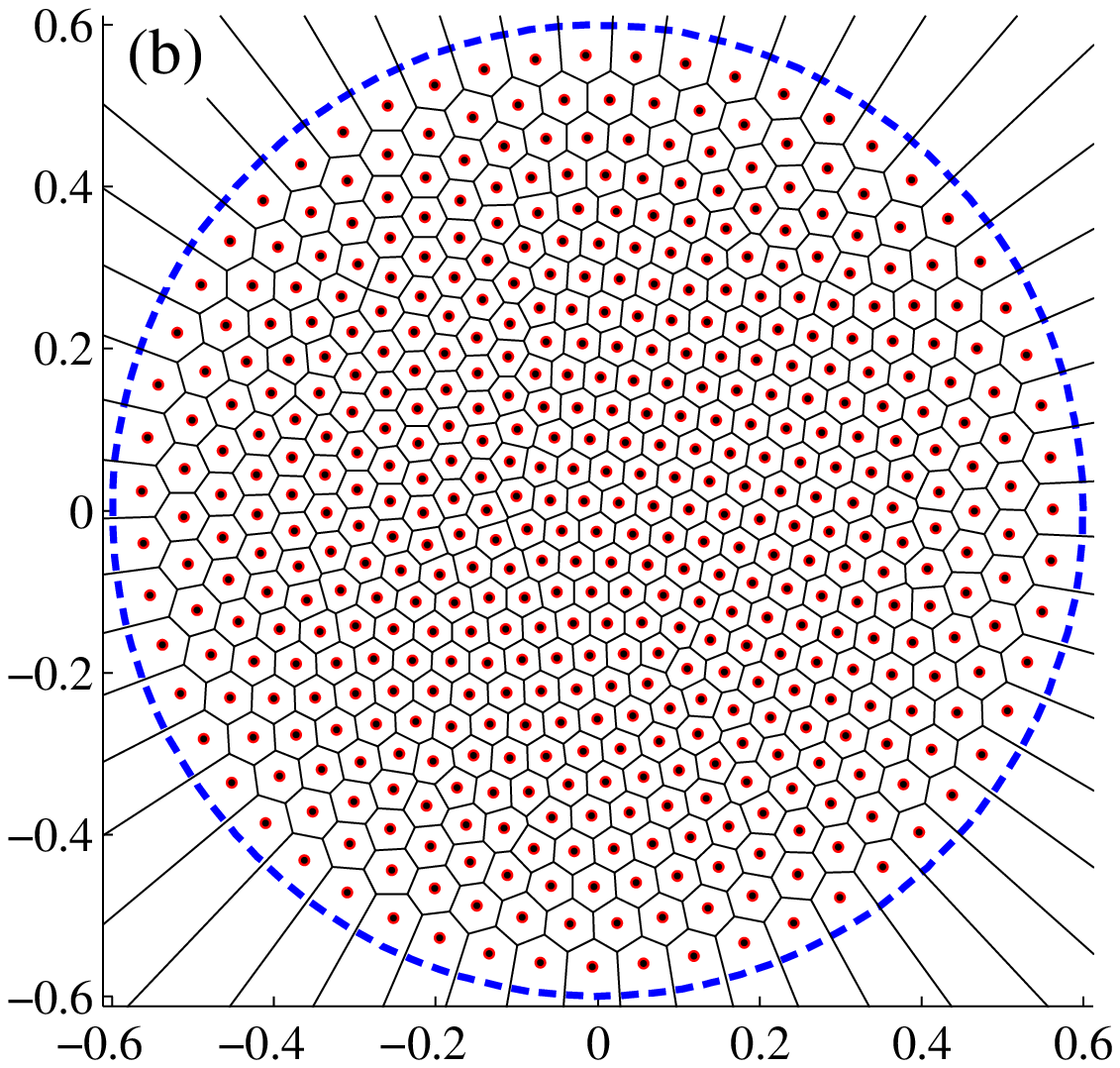}\\[2.0ex]
\includegraphics[width=0.43\textwidth]{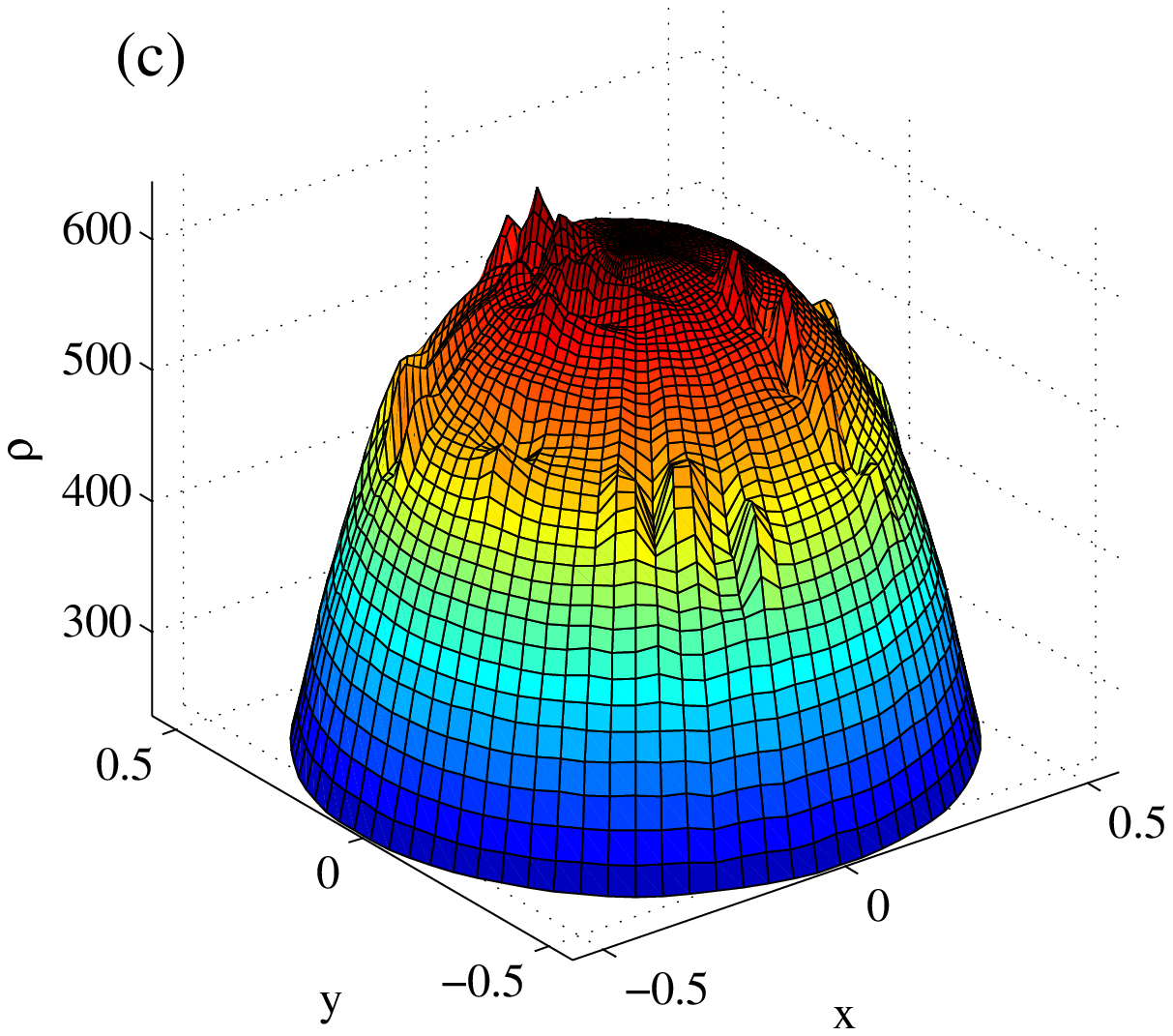}&~~~
\includegraphics[width=0.43\textwidth]{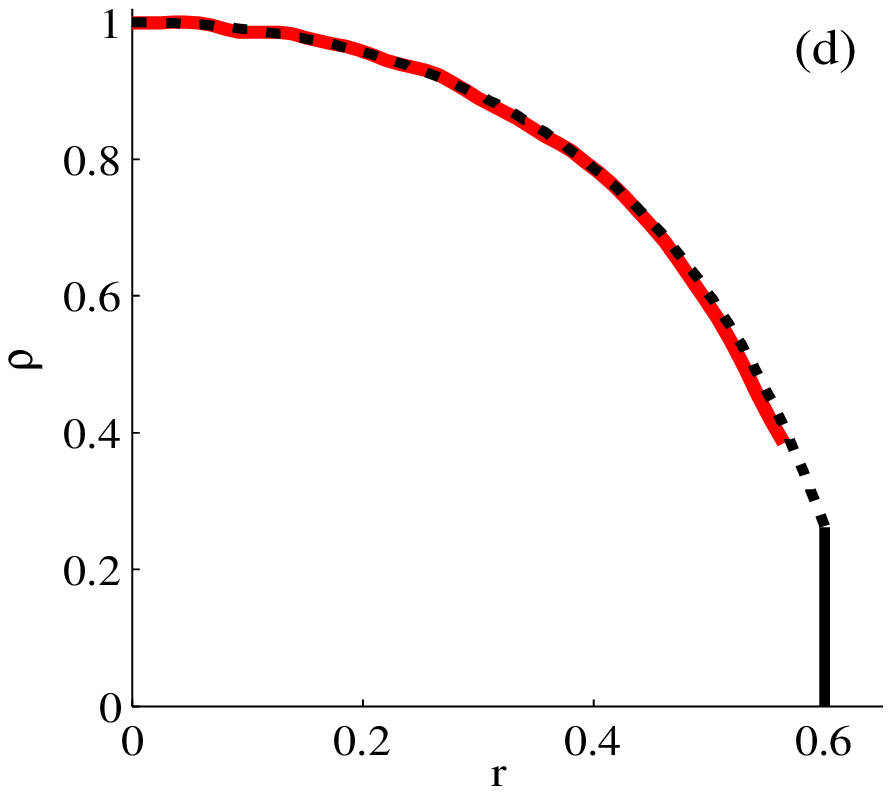}
\end{tabular}
\caption{(a) Stable
equilibrium of Eq.~(\ref{aggr}) with $f(r)$ as in Eq.~(\ref{f}). Parameter values are $N=500,
\omega=2.95139, a=1$ and $c=0.001$. The dashed circle is the asymptotic
boundary whose radius $R=0.6$ is the smaller solution to Eq.~(\ref{R}).
(b) Voronoi diagram used to compute the two-dimensional density distribution.
(c) The corresponding density distribution $\rho$ obtained by setting $\rho
(x_{j})=1/\mbox{area}_{j}$ and extrapolating, where $\mbox{area}_{j}$ is the
area of the Voronoi cell that contains $x_{j}$.
(d) Average of $\rho
(|x|)/\rho(0)$ as a function of $r=|x|$. The solid curve corresponds to the
numerical computation. The 
dashed curve is the formula (\ref{rho}) and the vertical line
is the boundary $r=R$. 
}%
\label{tkfig1}
\end{figure}
\end{empty}

Following, e.g., the discussion of Ref.~\cite{bernoff}, we coarse-grain by defining
the particle density to be%
\begin{equation}
\rho(x)=\sum_{k=1\ldots N}\delta(x-x_{k}),
\end{equation}
where $\delta(x)$ is the Dirac-delta function.
It is then straightforward to rewrite the aggregation Eq.~(\ref{aggr}) as
$\dot{x}_{j}=v(x_{j})$ where we define the continuum limit of the velocity as%
\[
v(x)\equiv\left(  f(r)-\omega\right)  x+c\int_{\mathbb{R}^{2}}
\frac{x-y}{\left\vert x-y\right\vert ^{2}}\rho\left(  y\right)  dy.
\]
The term $f(r)=a/(1-r^{2})$ is relevant, in particular, to the precessional
dynamics of interest in quasi-two-dimensional trapped BECs. Here $r=\left\vert
x\right\vert $ represents the radial variable and the density normalization
condition reads%
\[
\int_{\mathbb{R}^{2}}\rho(x)dx=N.
\]
In the limit of large $N$, conservation of mass then yields the
following continuity equation:
\begin{equation}
\rho_{t}+\nabla\cdot\left(  v\rho\right)  =0.
\end{equation}

Assuming in this large $N$ limit a radially symmetric density, we note that
for any smooth radial function $g(r),$ we have the following identity:%
\[
\int_{\mathbb{R}^{2}}\frac{x-y}{\left\vert x-y\right\vert ^{2}}g\left(  \left\vert
y\right\vert \right)  dy=x\frac{2\pi}{r^{2}}\int_{0}^{r}g(s)s\, ds,
\]
so that [with slight abuse of notation $\rho(y)=\rho(|y|)$],
\begin{equation}
v=\left(  f(r)-\omega+\frac{2\pi c}{r^{2}}\int_{0}^{r}\rho(s)sds\right)  x.
\label{v=}%
\end{equation}
Let $V(r)$ be the bracketed expression in Eq.~(\ref{v=})\ so that $v=V(r)x$ and
note that%
\[
\nabla\cdot\left(  v\rho\right)  =\nabla\cdot\left(  V(r)\rho x\right)
=\frac{1}{r}\left(  V\rho r^{2}\right)  _{r}%
\]
so that $\rho_{t}+\nabla\cdot\left(  v\rho\right)  =0$ becomes%
\begin{equation}
(r\rho)_{t}+\left(  Vr\rho r\right)  _{r}=0. \label{conrad}%
\end{equation}
Now define%
\[
u(r)=\int_{0}^{r}\rho(s)sds.
\]
Integrating Eq.~(\ref{conrad})\ we obtain%
\begin{equation}
u_{t}+Vru_{r}=0. \label{u}%
\end{equation}
Recall that we have
\begin{align}
Vr&=r\left(  f(r)-\omega\right)  +\frac{2\pi c}{r}\int_{0}^{r}\rho
(s)sds
\nonumber
\\
\nonumber
&=r\left(  f(r)-\omega\right)  +\frac{2\pi c}{r}u.
\end{align}
Thus we obtain the following characteristics for Eq.~(\ref{u})
\begin{align}
\frac{dr}{dt}&=r\left(  f(r)-\omega\right)  +\frac{2\pi c}{r}u,
\nonumber
\\[2.0ex]
\frac{du}{dt}&=0. \label{char}%
\end{align}
Now suppose that the initial density is radially symmetric and has finite
support of radius $R.$ Then we have:
\begin{equation}
u(R)=\int_{0}^{R}r\rho dr=\frac{N}{2\pi}; \label{urho}%
\end{equation}
the corresponding characteristic $r=R$ then evolves according to
\begin{equation}
\frac{dR}{dt}=R\left(  f(R)-\omega\right)  +\frac{cN}{R}. \label{Rode}%
\end{equation}
In particular at the steady state $t\rightarrow\infty$ and with
$f(R)=a/(1-R^{2}),$ we obtain the equation for the support radius,%
\begin{equation}
\omega=\frac{a}{1-R^{2}}+\frac{cN}{R^{2}}. \label{R}%
\end{equation}

\begin{empty}
\begin{figure}[tb]
\begin{tabular}{ccc}
$N=25$& $N=100$&$N=400$\\
\includegraphics[width=0.31\textwidth]{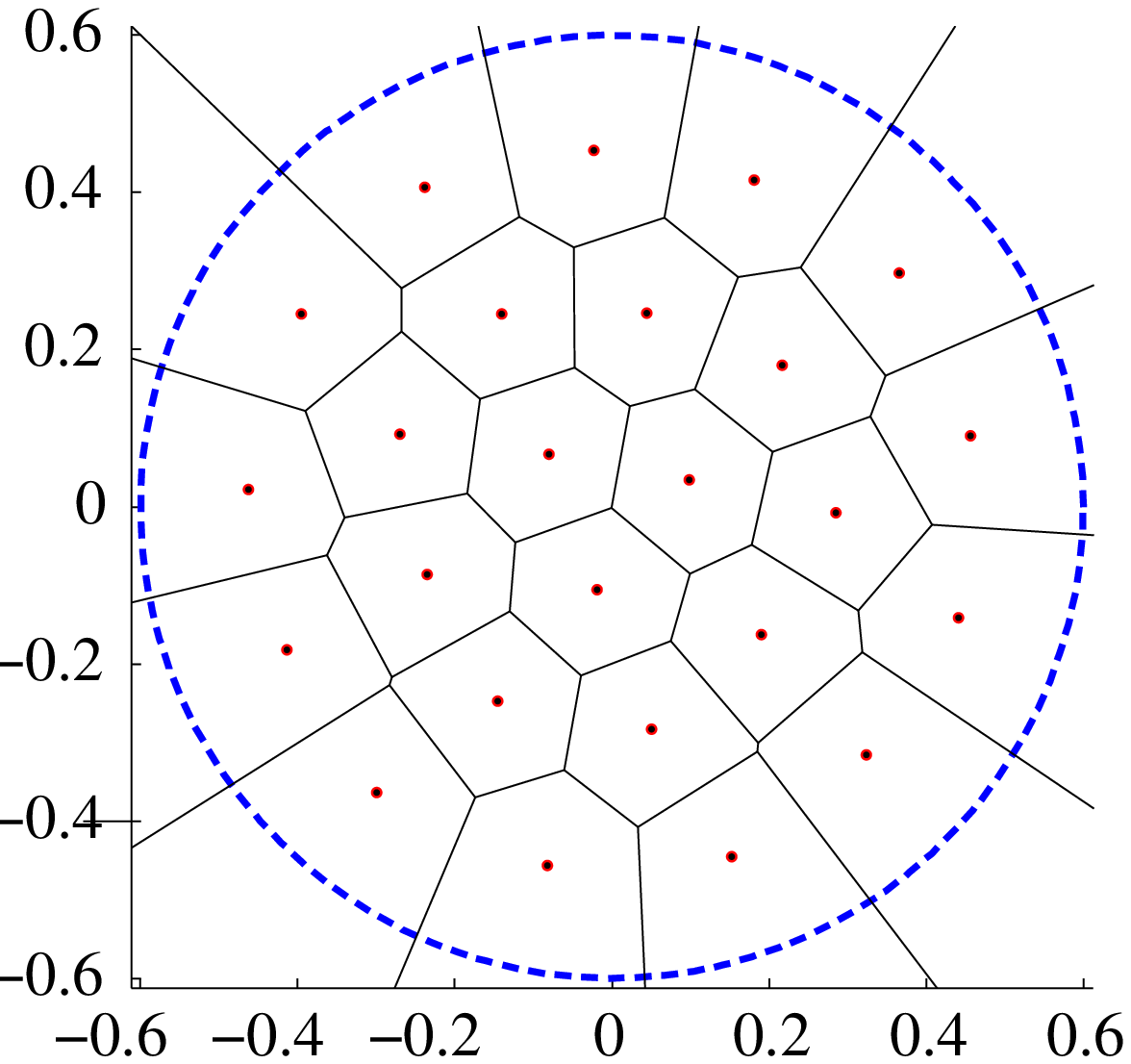}&
\includegraphics[width=0.31\textwidth]{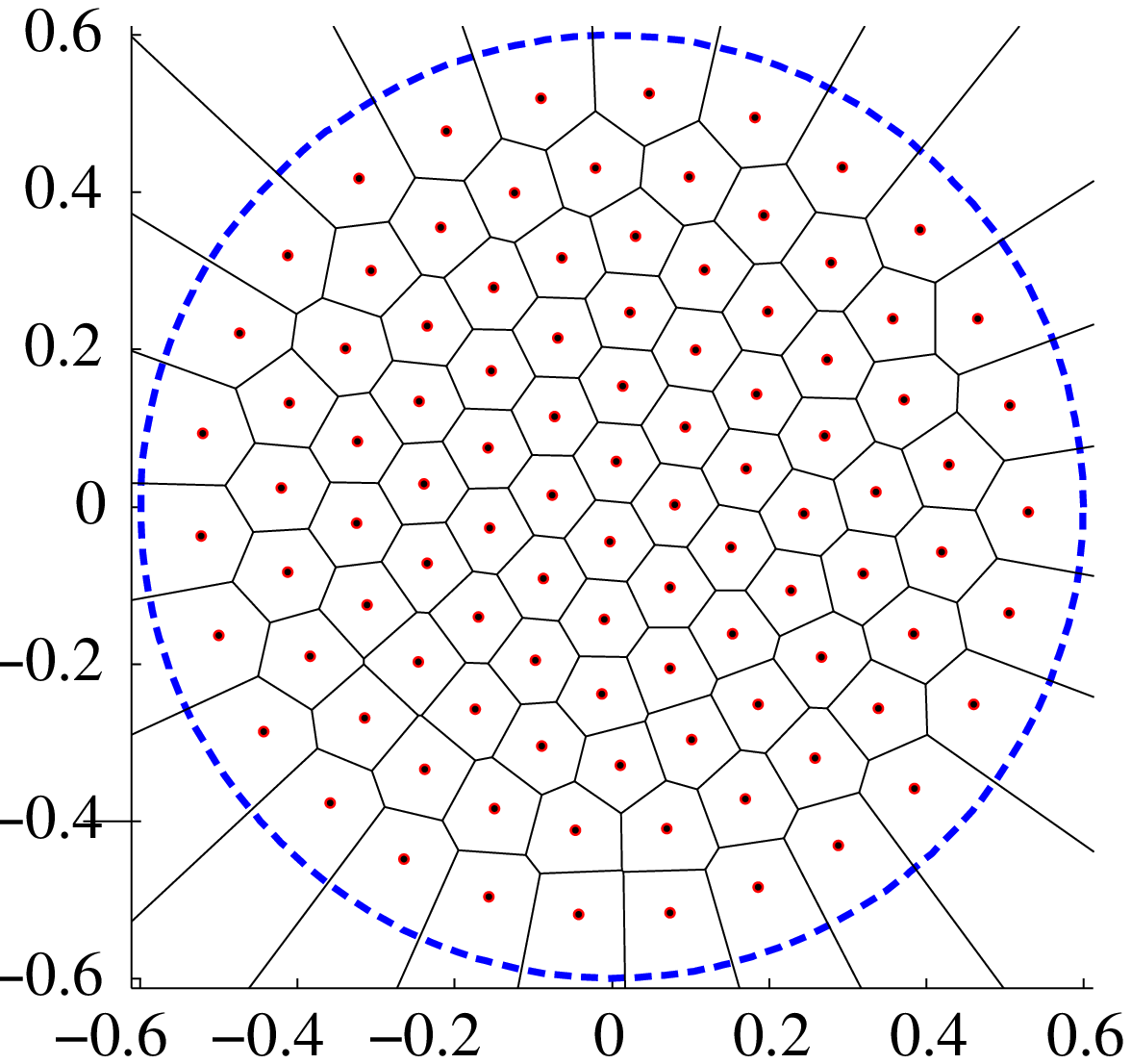}&
\includegraphics[width=0.31\textwidth]{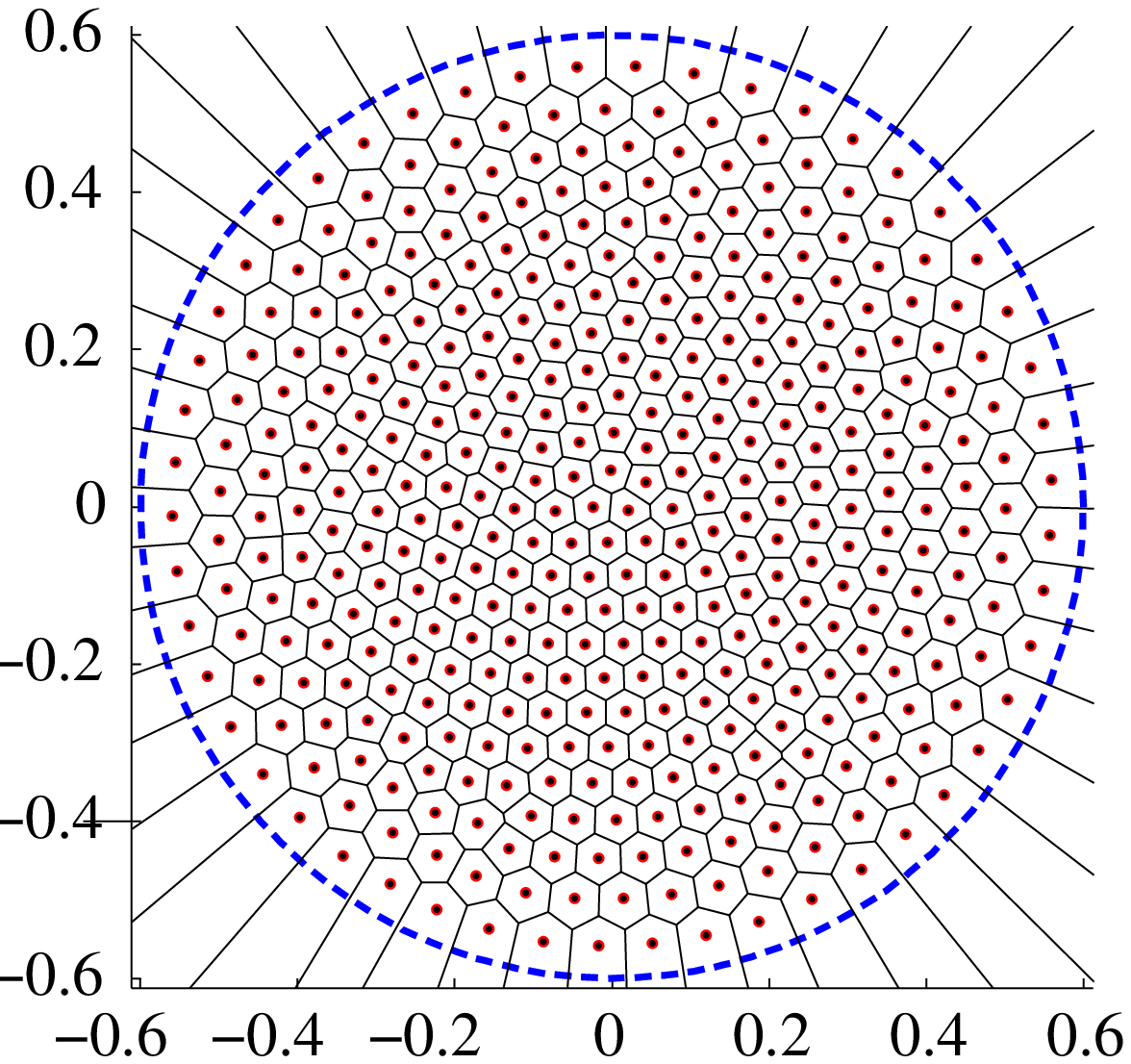}\\[2.0ex]
\includegraphics[width=0.31\textwidth]{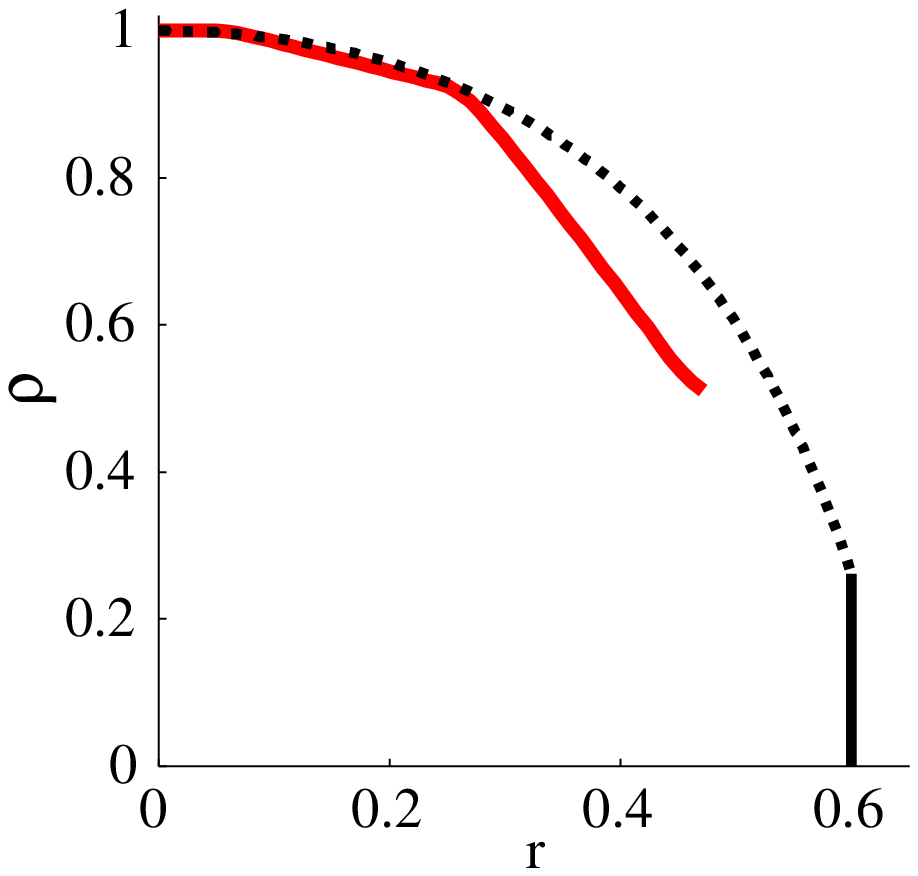}&
\includegraphics[width=0.31\textwidth]{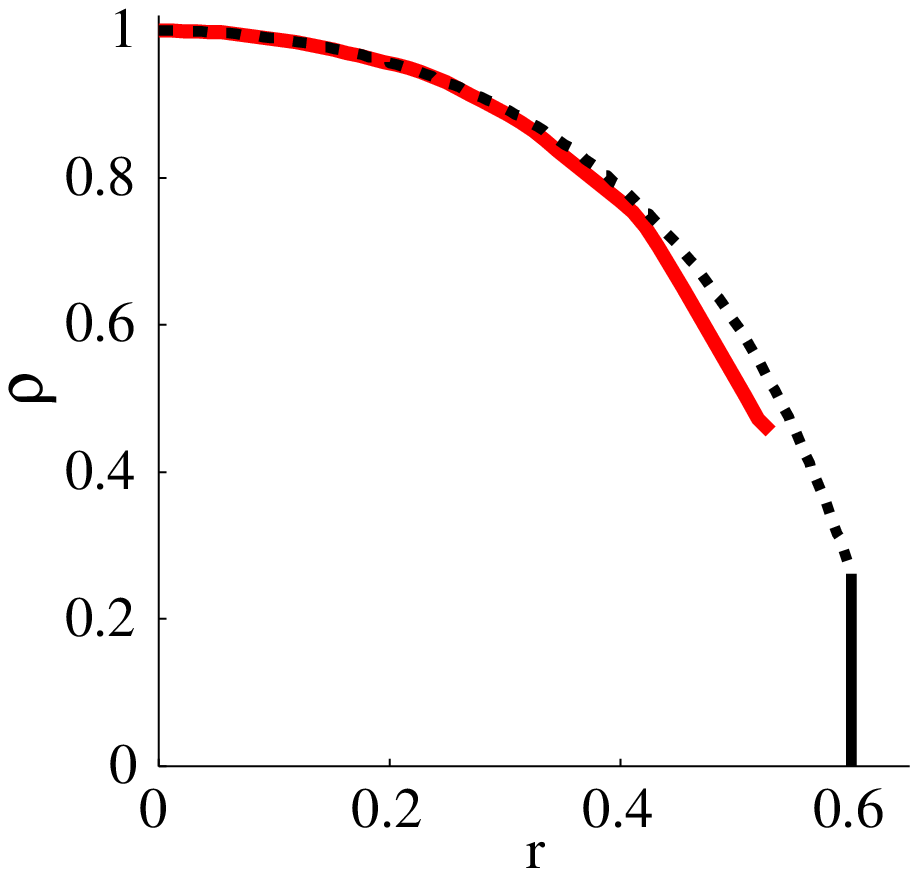}&
\includegraphics[width=0.31\textwidth]{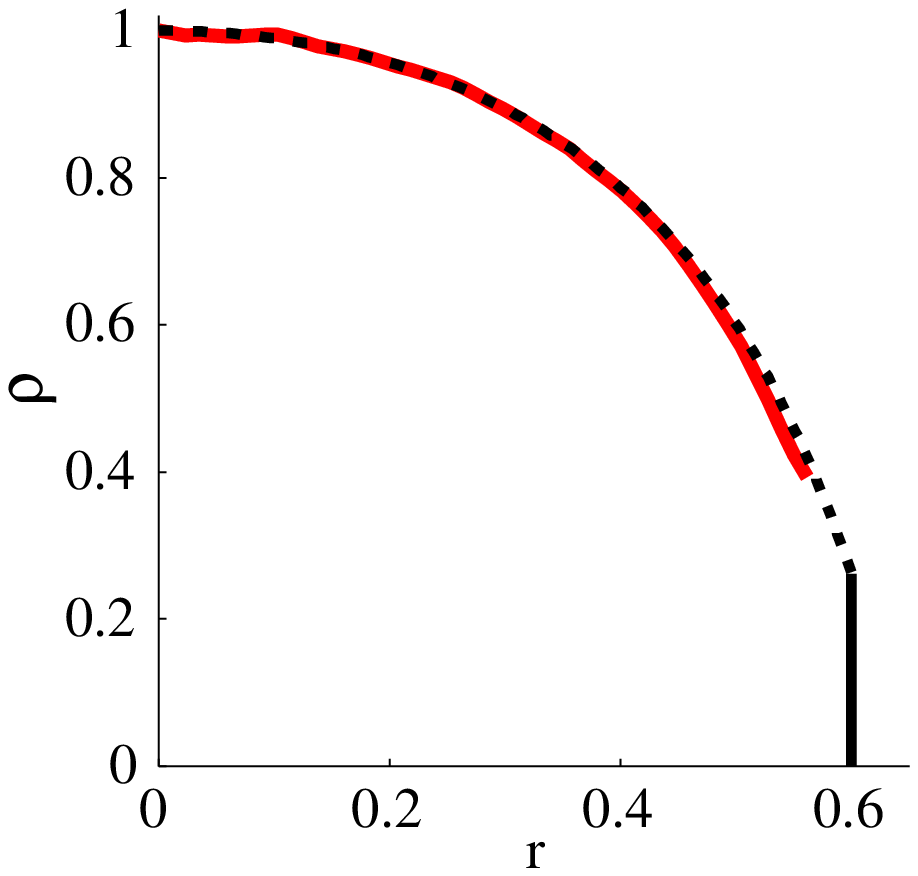}
\end{tabular}
\caption{Top row: stable
equilibrium of Eq.~(\ref{aggr}) with $f(r)$ as in Eq.~(\ref{f}), with
$N$ as shown in the titles and $c=0.5/N$,
$\omega=2.95139$, and $a=1$. The dashed circle is the asymptotic
boundary whose radius $R=0.6$ is the smaller solution to Eq.~(\ref{R}).
Bottom row: average of $\rho$
$(|x|)/\rho(0)$ as a function of $r=|x|$. The solid curve corresponds to the
numerical computation. The dashed curve is the formula (\ref{rho})
and the vertical line is the boundary $r=R$. }
\label{tkfig2}
\end{figure}
\end{empty}

From Eq.~(\ref{char}), at the steady state we have $u=-r^{2}\left(  f(r)-\omega
\right)  /(2\pi c)$. From Eq.~(\ref{urho}) we then obtain $\rho=-\frac{1}{2\pi
c}\frac{1}{r}\frac{\partial}{\partial r}\left(  r^{2}\left[  f(r)-\omega
\right]  \right)  $ or%
\begin{equation}
\rho=\frac{1}{\pi c}\left(  \omega-\frac{a}{\left(  1-r^{2}\right)  ^{2}%
}\right).
  \label{rho}%
\end{equation}
Note that for fixed $\omega$, Eq.~(\ref{R}) has either zero or (one
at the critical point or) two solutions for $R.$ If it has two solutions
$R_{-}<R_{+},$ then $R_{-}$ is stable and $R_{+}$ is unstable, as is easily
deduced from Eq.~(\ref{Rode}). The threshold occurs by setting $\partial
\omega/\partial R=0$ to obtain
\begin{align}
\omega_{c}=\left(  \sqrt{a}+\sqrt{cN}\right)  ^{2},
\nonumber
\qquad
R_{c}^{2}=\frac{\sqrt{cN}}{\sqrt{a}+\sqrt{cN}}.
\end{align}
Thus two solutions exist when $\omega>\omega_{c}:$ the one with smaller
support is stable and the one with bigger support is unstable. Interestingly,
the density $\rho(R)$ vanishes precisely at $R=R_{c}$. Hence, among the two
solutions only the stable one with $R_{-}<R_{c}$ is physically relevant, while
the unstable one with $R_{+}>R_{c}$ cannot be computationally obtained (and
presumably physically observed since it would involve negative
vortex densities for $R>R_c$). In order to compute the steady-state
distribution of the vortices, we first evolved Eq.~(\ref{aggr}) until it settled
to an equilibrium state [Fig.~\ref{tkfig1}(a)]. We then computed the Voronoi
tessellation of the plane using the Matlab function \texttt{voronoi}
[Fig.~\ref{tkfig1}(b)]. This tessellation assigns to each vortex $x_{j}$ a
region (Voronoi cell)\ which consists of all points in the plane that are
closer to $x_{j}$ than to any other vortex. We then approximated the density
distribution at $x_{j}$ by $\rho\left(  x_{j}\right)  =1/{\rm area}_{j}$ where
${\rm area}_{j}$ is the area of the Voronoi cell associated to $x_{j}.$ The
resulting distribution (extrapolated linearly between the points) is
plotted in Fig.~\ref{tkfig1}(c). Finally, in Fig.~\ref{tkfig1}(d) we plot
the radial density $\rho(r),$ which we computed by taking the average of
$\rho(x)$ along $\left\vert x\right\vert =r$. We compare this to the
distribution from the analytical expression of Eq.~(\ref{rho}). There is an
excellent agreement between the two corroborating the value of our theoretical prediction.
Figure~\ref{tkfig2} shows that this agreement persists for smaller 
values of $N$ (e.g., $N=25$) as well, although naturally it becomes 
progressively worse as $N$ is decreasing.

In a future work, we plan to study the stability of the steady state
(\ref{rho}). Numerical computations of Eq.~(\ref{aggr}) show that solution (\ref{rho}) is
indeed a stable equilibrium for the aggregation model. The corresponding
relative equilibrium of the BEC\ model (\ref{bec}) is then neutrally stable
and has vibrational or the so-called Tkachenko modes \cite{eng13a,tkachenko2,tkachenko3}.
We plan to extend the techniques in this section to compute the
vibrational modes in the continuum limit of large $N$.

\section{Conclusions and Future Challenges}
\label{sec:conclu}

In summary, in the present work we have revisited two opposite limits of the
quasi-two-dimensional co-rotating vortex dynamics in Bose-Einstein
condensates. Motivated by the recent success of particle models in capturing
experimental features of both the counter- and co-rotating vortex case, we
have attempted to examine in detail (fully analytically, wherever possible)
both the small $N$ and the large $N$ limit of such $N$-vortex configurations.
In the former case, we obtained vortex configurations 
in the form of polygonal rings. We
generalized the classical result of Ref.~\cite{havel} unveiling that the ring is
generically unstable for $N\geq7$ in the case of monotonic precessional
frequency dependence on the distance from the trap center. Moreover, we showed
that the critical contribution of the precessional term creates the potential
for {stable asymmetric, as well as other configurations} even for
$N=2,\dots,6$, for sufficiently high angular momentum. In that light, we also
mentioned in passing the $N+1$ vortex configuration, whose stability is
analyzed in Appendix \ref{appII}. The opposite limit of large $N$ 
is quite interesting
in its own right. Since polygonal configurations are already highly unstable
for sufficiently large $N$, a fundamentally different distribution is expected
for large $N$. This distribution was identified in a radial form, by looking
at the corresponding continuum equation and was corroborated numerically.
An
ongoing collaboration with the group that has made critical earlier 
experimental contributions in this theme (see 
Refs.~\cite{freilich10,navar13}) suggests the
feasibility of looking at controllably small numbers of $N$ (up to $11$)
as well as at the regime of
large $N$ regime experimentally.

The results in Sec.~\ref{sec:continuum} generalize 
a similar computation 
for classical vortex dynamics \cite{Kolokolnikov:2013}. 
The methods used here and in Ref.~\cite{Kolokolnikov:2013}
are borrowed from the literature on biological 
swarming, see, e.g., Refs.~\cite{fetecau, bernoff}.
Similar techniques were also recently used to study predator-swarm dynamics \cite{pred-swarm}. It is hoped
that further developments in the swarming literature will help to shed light on 
the behavior of BEC (and in particular, the stability of vortex lattices) and vice-versa.

There are numerous directions in which we foresee that this activity can be
extended. On the one hand, it would be particularly interesting (since the
experimental possibilities reported in Ref.~\cite{navar13} allow the ``dialing 
in''
of different numbers of vortices, e.g., between 1 and 11) to explore further the case
of intermediate-size clusters i.e., between $N=5$ and $N=11$. There,
identifying the potential $N$-vortex ring polygons, or that of $N+1$ rings or
the examination of different ground state configurations would be relevant to
examine. On the other hand, in the case of large $N$, our preliminary
computations (via fixed point iterations of a Newton scheme) reveal a large
number of excited state configurations. It will be interesting to explore in
future studies whether these are generically unstable or whether additional
dynamically stable large $N$ limits could, in principle be accessible as well.
Furthermore, examinations of multi-component (e.g., pseudo-spinor) settings
in Refs.~\cite{eng13a,eng13b}, of potentials of different symmetry (such as square optical
lattices, which can again induce structural phase transitions~\cite{leslie})
motivate analogous considerations/extensions at the level of our particle
model. 

While the mean-field theory is successful at predicting the large-scale vortex
density distribution, it does not capture the fine structure of the BEC lattice
itself; see, e.g., Ref.~\cite{cipriani} where different lattices 
and where their dynamics  and internal 
(Tkachenko) modes were observed~\cite{simula1}.
%
The point vortex BEC model (\ref{bec}) is an approximation to the full 
system more accurately described by a three-dimensional Gross-Pitaevskii 
model, while neglecting the vortex core structure. 
The three-dimensionality can lead to more complex configurations such 
as ``Olympic rings'', see, e.g., Refs.~\cite{simula1,simula2}.
It would be
interesting to see if the techniques of this paper could also be applied to such configurations.
Finally, the examination of trapped, interacting vortex rings in 
three dimensions
both in the context of few~\cite{brand1,brand2,komineas,konstantinov} and in
that of many such rings would be a broad direction of considerable importance
for future studies.

\section{Acknowledgements}
We thank the anonymous referees for their valuable comments, 
which significantly improved the paper.
T.K.~was supported by NSERC grant no.~47050.
P.G.K.~and R.C.G.~gratefully acknowledge the support of NSF-DMS-0806762
and NSF-DMS-1312856 and NSF-CMMI-1000337.

\appendix
\section{Appendix}

\subsection{
\label{appI}
Relation between stability of aggregation and BEC equations}

We prove Theorem \ref{thm:stab} here. Linearize Eq.~(\ref{aggr}) around the steady
state $x_{j}(t)=\xi_{j}$ by using $x_{j}(t)=\xi_{j}+\eta_{j}(t)$ with
$|\eta_{j}|\ll1.$ We then obtain the system%
\begin{equation}
\dot{\eta}=D\eta+L\bar{\eta}.
\label{recycles}%
\end{equation}
Here $\eta=\left(  \eta_{1}\ldots\eta_{N}\right)  ^{T}$ is the perturbation
vector; overbar denotes the complex conjugate; $L$ is a symmetric complex
matrix whose entries are
\[
L_{jk}=\left\{
\begin{array}
[c]{c}%
\dfrac{c}{\left(  \overline{\xi_{j}-\xi_{k}}\right)  ^{2}},\ \ \ \ \ j\neq k,
\\[4.0ex]
f^{\prime}\left(  \left\vert \xi_{j}\right\vert \right)  \dfrac{\xi_{j}^{2}%
}{2\left\vert \xi_{j}\right\vert }-
{\displaystyle\sum\limits_{k\neq j}}
\dfrac{c}{\left(  \overline{\xi_{j}-\xi_{k}}\right)  ^{2}},\ \ \ \ \ \ \ j=k,
\end{array}
\right.
\]
and $D$ is a diagonal real matrix whose entries are%
\begin{equation}
\label{matrixD}
D_{jj}=f\left(  \left\vert \xi_{j}\right\vert \right)  +f^{\prime}\left(
\left\vert \xi_{j}\right\vert \right)  \dfrac{\left\vert \xi_{j}\right\vert
}{2}-\omega.
\end{equation}
By taking the complex conjugate of Eq.~(\ref{recycles}) we obtain a 
closed system of $2N$ ODEs given by
\begin{equation}
\label{linsys}
\left\{
\begin{array}{rcl}
\partial_{t}\eta&=&D\eta+L\bar{\eta},
\\[2.0ex]
\partial_{t}\bar{\eta}&=&D\bar{\eta}+\bar{L}\eta.
\end{array}
\right.
\end{equation}
Linearizing around the steady state equilibrium, we find that the eigenvalues
of the zero equilibrium of Eq.~(\ref{linsys}) are given by the matrix%
\[
A=\left[
\begin{array}
[c]{cc}%
D & L\\
\bar{L} & D
\end{array}
\right]  .
\]
Performing a similar analysis the relative equilibrium $z_{j}(t)=e^{i\omega
t}\xi_{j}$ of Eq.~(\ref{bec}), we find that its eigenvalues are given by the
matrix%
\[
B=i
J
A=\left[
\begin{array}
[c]{cc}%
iD & iL\\
-i\bar{L} & -iD
\end{array}
\right],
\]
where
\begin{equation}
\nonumber
J=\left[
\begin{array}[c]{cc}%
I & 0\\
0 & -I
\end{array}
\right].
\end{equation}
Next, we show that if all eigenvalues of $A$ are strictly negative, then all
eigenvalues of $B$ are purely imaginary. Since $A$ is Hermitian, we may write
$A=UE\bar{U}^{T}$ where $E$ is a diagonal matrix whose diagonal entries are
the eigenvalues of $A,$ and $U$ is unitary. Assume that all eigenvalues of $A$
are negative. Then we can write $E=-Q^{2}$ where $Q$ is a real diagonal
matrix, so that $B=-iJUQQ\bar{U}^{T}$.
Note that in general, the spectrum of matrices $M_{1}M_{2}$ and
$M_{2}M_{1}$ is the same, so that $B$ has the same spectrum as the matrix
$-iQ\bar{U}^{T}JUQ$ whose eigenvalues are purely imaginary since
$Q\bar{U}^{T}JUQ$ is Hermitian.

To show the converse, note that under either conditions (i) or (ii) of Theorem
\ref{thm:stab}, the matrix $D$ given by Eq.~(\ref{matrixD})
is a multiple of identity so we may write
$D=dI$ where $d$ is a constant. In this case, the eigenvalues of $A$ are given
by $\lambda_{A}=d\pm\sqrt{\epsilon}$ where $\epsilon\in\mathbb{R}^{+}$ 
is an eigenvalue of $L\bar{L};$ whereas the eigenvalues of $B$ are given
by $\lambda_{B}=\pm\sqrt{\epsilon-d^{2}}.$ It follows that $\lambda_{B}$ is
purely imaginary if and only if $\epsilon<d^{2},$ which is if and only if
$\lambda_{A}=d\pm\sqrt{\epsilon}<0.$ $\ \ \blacksquare$


\subsection{
\label{appII}
N+1 state: ring solution with a vortex at the center}

Here we prove Theorem \ref{thm:N+1}. Similar to the ring steady state, we
consider the relative equilibrium of the aggregation model with $N+1$
vortices;\ $N$ on the ring and one at the center. As in
Ref.~\cite{cabral2000stability}, we will actually consider a slightly more general
problem where the central vortex has weight $b$ whereas other vortices have
weight $c$; Theorem \ref{thm:N+1} will follow by taking $b=c.$ The starting
point is
\begin{equation}
\nonumber
\left\{
\begin{array}{rcl}
\displaystyle
\dot{z}_{j}  &=&
\displaystyle
\left(  f\left(  \left\vert z_{j}\right\vert \right)-\omega\right)  z_{j}
+c\sum_{k\neq j}\frac{z_{j}-z_{k}}{\left\vert z_{j}-z_{k}\right\vert ^{2}}
+b\frac{z_{j}-z_{N+1}}{\left\vert z_{j}-z_{N+1}%
\right\vert ^{2}},\ \ \ j=1\ldots N,
\\[5.0ex]
\dot{z}_{N+1}   &=&
\displaystyle
\left(  f\left(  \left\vert z_{N+1}\right\vert \right)
-\omega\right)  z_{N+1}
+c\sum_{k\neq j}^{N}\frac{z_{N+1}-z_{k}}{\left\vert
z_{N+1}-z_{k}\right\vert ^{2}}.
\end{array}
\right.
\end{equation}
As before, we make the ansatz
\begin{equation}
\nonumber
\left\{
\begin{array}{rcl}
z_{j}(t)&=&R\,\exp\left(  \frac{2\pi i}{N}j\right),\ \ j=1\ldots N,
\\[2.0ex]
z_{N+1}(t)&=&0.
\end{array}
\right.
\end{equation}
Then $R$ satisfies
\begin{equation}
\omega=f\left(  R\right)  +\frac{c\left(  N-1\right)  }{2R^{2}}+\frac{b}%
{R^{2}}; \label{funnybunny}%
\end{equation}
setting $b=c$ recovers formula (\ref{R2}).

Next we consider perturbations to the $N+1$ vortex configurations.
As before we perturb the steady state as
\begin{align*}
x_{k}(t)  &  =R\,\xi^{k}\left(  1+h_{k}(t)\right)  ,\ \ \ |h_{k}|\ll 1,
\ \ k=1\ldots N
\\[2.0ex]
x_{N+1}(t)  &  =R\,h_{N+1}(t),
\end{align*}
where we defined%
\[
\xi\equiv\exp\left(  2\pi i/N\right),
\]
to obtain
\begin{align}
&
\displaystyle
\frac{dh_{j}}{dt}
=
\displaystyle
\left(  f^{\prime}(R)\frac{R}{2}+f(R)-\omega\right)
h_{j}+f^{\prime}(R)\frac{R}{2}\bar{h}_{j}
+c\sum_{k\neq j}^{N}\frac{\xi
^{k-j}\bar{h}_{j}-\bar{h}_{k}}{4R^{2}\sin^{2}\left(  \frac{\pi\left(
k-j\right)  }{N}\right)  }-b\frac{\bar{h}_{j}-\xi^{j}\bar{h}_{N+1}}{R^{2}},
\nonumber
\\[0.0ex]
\nonumber
&
\displaystyle
\frac{dh_{N+1}}{dt}
=
\displaystyle
\left(  f(0)-\omega\right)  h_{N+1}+c\sum_{k=1}%
^{N}\frac{\xi^{k}\bar{h}_{k}}{R^{2}}.
\end{align}
The solution decomposes into a product of two subspaces:

\noindent
\textbf{Subspace 1:} Use the ansatz
\begin{equation}
\nonumber
h_{j}(t)=\xi_{+}(t)\xi^{mj}+\xi_{-}(t)\xi^{-mj},\ m\in\mathbb{N}
,\ \ j=1\ldots N;\ \ \ \ \ \ h_{N+1}=0,
\end{equation}
and collecting like terms in $e^{im2\pi j/N}$ and $e^{-im2\pi j/N},$ the
system (\ref{h})\ decouples into a sequence of $2\times 2$ subproblems
\begin{align}
\xi_{+}^{\prime}  =&\left(  f^{\prime}(R)\frac{R}{2}+f(R)-\omega\right)
\xi_{+}+\left(  f^{\prime}(R)\frac{R}{2}-\frac{b}{R^{2}}\right)  \bar{\xi}_{-}
\nonumber
+\bar{\xi}_{-}c\sum_{k,k\neq j}\frac{\xi^{k-j}-\xi^{m(k-j)}}{4R^{2}%
\sin^{2}\left(  \pi\left(  k-j\right)  /N\right)  },
\nonumber
\\[2.0ex]
\nonumber
\xi_{-}^{\prime}  =&\left(  f^{\prime}(R)\frac{R}{2}+f(R)-\omega\right)
\xi_{-}+\left(  f^{\prime}(R)\frac{R}{2}-\frac{b}{R^{2}}\right)  \bar{\xi}_{+}
\nonumber
+\bar{\xi}_{+}c\sum_{k,k\neq j}\frac{\xi^{k-j}-\xi^{-m(k-j)}}{4R^{2}%
\sin^{2}\left(  \pi\left(  k-j\right)  /N\right)  },
\end{align}
and, as previously, we obtain
\[
\lambda_{\pm}(m)=f^{\prime}(R)\frac{R}{2}+f(R)-\omega\pm\left(  f^{\prime
}(R)\frac{R}{2}-\frac{b}{R^{2}}+\frac{c}{2R^{2}}\left(  m-1\right)  \left(
N-m-1\right)  \right)  ,\ \ m=0\ldots N-1.
\]
Using Eq.~(\ref{funnybunny}) yields
\[
\lambda_{+}(m)=f^{\prime}(R)R+\frac{c}{2R^{2}}\left\{  \left(  m-1\right)
\left(  N-m-1\right)  -(N-1)\right\}  -\frac{2b}{R^{2}},
\]
with $\lambda_{-}(m)\leq0$ for all $m.$ As in Theorem \ref{thm:stabring}, this
expression is maximized when $m=\left\lfloor N/2\right\rfloor$.
Setting $b=c$ recovers Eq.~(\ref{lampstar}).

\noindent
\textbf{Subspace 2:} we use the ansatz%
\begin{equation}
\nonumber
h_{j}(t)=\xi_{+}(t)\xi^{j}+\xi_{-}(t)\xi^{-j},\ \ \ j=1\ldots
N;\ \ \ \ \ \ h_{N+1}=\eta(t),
\end{equation}
which yields%
\begin{align}
\nonumber
\xi_{+}^{\prime}  &  =\left(  f^{\prime}(R)\frac{R}{2}+f(R)-\omega\right)
\xi_{+}+\left(  f^{\prime}(R)\frac{R}{2}-\frac{b}{R^{2}}\right)  \bar{\xi}%
_{-}+b\frac{\bar{\eta}}{R^{2}},
\\[2.0ex]
\xi_{-}^{\prime}  &  =\left(  f^{\prime}(R)\frac{R}{2}+f(R)-\omega\right)
\xi_{-}+\left(  f^{\prime}(R)\frac{R}{2}-\frac{b}{R^{2}}\right)  \bar{\xi}%
_{+},
\nonumber
\\
\nonumber
\frac{d\eta}{dt}  &  =\left(  f(0)-\omega\right)  \eta+c\sum_{k=1}^{N}%
\frac{\bar{\xi}_{+}}{R^{2}},
\end{align}
or%

\[
\nonumber
\left(
\begin{array}
[c]{c}%
\xi_{+}^{\prime}
\\[2.0ex]
\bar{\xi}_{-}^{\prime}
\\[2.0ex]
\bar{\eta}^{\prime}%
\end{array}
\right)  =\left(
\begin{array}
[c]{ccc}%
f^{\prime}(R)\frac{R}{2}+f(R)-\omega & f^{\prime}(R)\frac{R}{2}-\frac{b}%
{R^{2}} & \frac{b}{R^{2}}
\\[2.0ex]
f^{\prime}(R)\frac{R}{2}-\frac{b}{R^{2}} & f^{\prime}(R)\frac{R}%
{2}+f(R)-\omega & 0
\\[2.0ex]
\frac{cN}{R^{2}} & 0 & \left(  f(0)-\omega\right)
\end{array}
\right)  \left(
\begin{array}
[c]{c}%
\xi_{+}
\\[2.0ex]
\bar{\xi}_{-}
\\[2.0ex]
\bar{\eta}%
\end{array}
\right)  .
\]
Substituting $b=c$ into the matrix above yields Eq.~(\ref{M0}). $\blacksquare$


\end{document}